# Lateral diffusion in 2 µm InGaAs/GaAsSb superlattice planar diodes using atomic layer deposition of ZnO

Manisha Muduli,[1] Nathan Gajaowski,[1] Hyemin Jung,[1] Neha Nooman,[1] Bhupesh Bhardwaj,[2] Mariah Schwartz,[1] Seunghyun Lee,[1] and Sanjay Krishna[1]

[1])*Electrical and Computer Engineering, The Ohio State University, OH, USA.*

[2])*Indian Institute of Technology, Bombay, Mumbai, MH, India.*

(*Electronic mail: krishna.53@osu.edu)

(Dated: 20 November 2023)

Abstract: Avalanche photodiodes used for greenhouse gas sensing often use a mesa-structure that suffers from high surface leakage currents and edge breakdown. In this paper, we report 2-µm InGaAs/GaAsSb superlattice (SL) based planar PIN diodes to eliminate the challenges posed by conventional mesa diodes. An alternate way to fabricate planar diodes using atomic layer deposited ZnO was explored and the effect of the diffusion process on the superlattice was studied using X-ray diffraction. The optimum diffusion conditions were then used to make planar PIN diodes. The diffused Zn concentration was measured to be approximately $10^{20} cm^{-3}$ with a diffusion depth of 50 nm and a lateral diffusion ranging between 18 µm to 30 µm. A background doping of $5.8 \times 10^{14} cm^{-3}$ for the UID layer was determined by analyzing the capacitance-voltage measurements of the superlattice PIN diodes. The room temperature dark current for a device with a designed diameter of 30 µm is 1 µA at -2V. The quantum efficiency of the diode with a designed diameter of 200 µm was obtained to be 11.11% at 2 µm illumination. Further optimization of this diffusion process may lead to a rapid, manufacturable, and cost-effective method of developing planar diodes.

Greenhouse gas sensing systems often use avalanche photodiodes (APDs) or PIN diodes operating between 2 µm to 2.4 µm for monitoring gases such as carbon dioxide ($CO_2$), carbon monoxide (CO), and methane ($CH_4$)[1,2].

$In_{0.53}Ga_{0.47}As/GaAs_{0.51}Sb_{0.49}$ superlattice (hereafter, InGaAs/GaAsSb SL) absorbs in the wavelength range of 2 µm, which makes it a good absorber for a greenhouse gas detector. Most detectors reported in the literature have a mesa structure[3–5]. Fabrication of mesa structure requires etching of the epitaxially grown material to isolate the detectors. The process of etching causes abrupt breaks in the periodicity of the material. This leads to the formation of impure bonds on the surface that cause Fermi-level pinning, increasing the surface leakage current[6–8]. Reducing the surface leakage current requires additional passivation steps, making the fabrication of the mesa structures complex and time-consuming[9–11].

Planar structured diodes use diffusion or implantation to define the detector pixel, which completely eliminates the need for etching. Thus, planar diodes reduce the number of fabrication steps, as well as reduce the surface effects that arise in mesa structures due to the etching process[6]. There are several ways to fabricate planar diodes, of which the commonly used methods are diffusion using MOCVD[6,12], ionimplantation[13,14] and thermal diffusion[15,16]. MOCVD can be difficult and expensive; ion-implantation can damage the epitaxially grown material[13], and thermal diffusion can be a complex process involving ampuoles[17]. To solve these challenges, we propose an alternate method for diffusing Zn using atomic layer deposition (ALD) of ZnO[18,19]. ALD can be used to deposit a very thin layer of ZnO at relatively low temperatures, which reduces the cost of fabrication and prevents any possible thermal damage. Additionally, the thin film of ZnO acts both as a surface passivant and a source for Zn dopant[19], eliminating any need for further surface passivation.

Material growth and Zn-diffusion: Unintentionally doped (UID) InGaAs/GaAsSb SL with UID InGaAs/n+ InGaAs/n++ InGaAs layers were grown on a lattice-matched InP substrate (001) using molecular beam epitaxy (MBE). Additional details about the material can be found in the reference[9,18].

ZnO films were grown on the UID InGaAs/GaAsSb SL layer using atomic layer deposition (ALD). The ALD precursor sources were diethyl zinc and water, and the growth temperature was 150 °C. The epitaxially grown stack, along with the grown ZnO thin film, is then put in a thermal furnace at a temperature of 450 °C for 1 hour in a forming gas atmosphere. The hydrogen in the forming gas reacts with the oxygen in the ZnO, breaking the bond between the zinc and oxygen atoms, which promotes zinc diffusion into the superlattice (equation 1). After the diffusion is complete, ZnO is removed using HCl(37%):H$_2$O [1:10]. The general diffusion process and the cross-section of the material stack after diffusion are shown in FIG. 1 (a).

$$ZnO + H_2 + N_2 \rightarrow Zn^{2+} + H_2O + N_2 \quad (1)$$

Three main factors that can affect diffusion are: 1) thickness of the ZnO thin film, 2) diffusion time in the furnace, and 3) diffusion temperature. To study the effect of ZnO thickness, four different thicknesses of ZnO were grown on the SL: 5 nm, 10 nm, 20 nm, and 30 nm. The different conditions reported in this paper are mentioned in TABLE I. Based on the previous report, diffusion time and temperature were chosen as 1 hour and 450 °C respectively[18]. The effect of the grown ZnO thickness, diffusion time and temperature was then compared with an as-grown sample. Fabrication of planar PIN diodes: For the device fabrication, we optimized the thickness of ALD-grown ZnO to 20 nm (FIG. A 1).

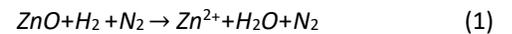

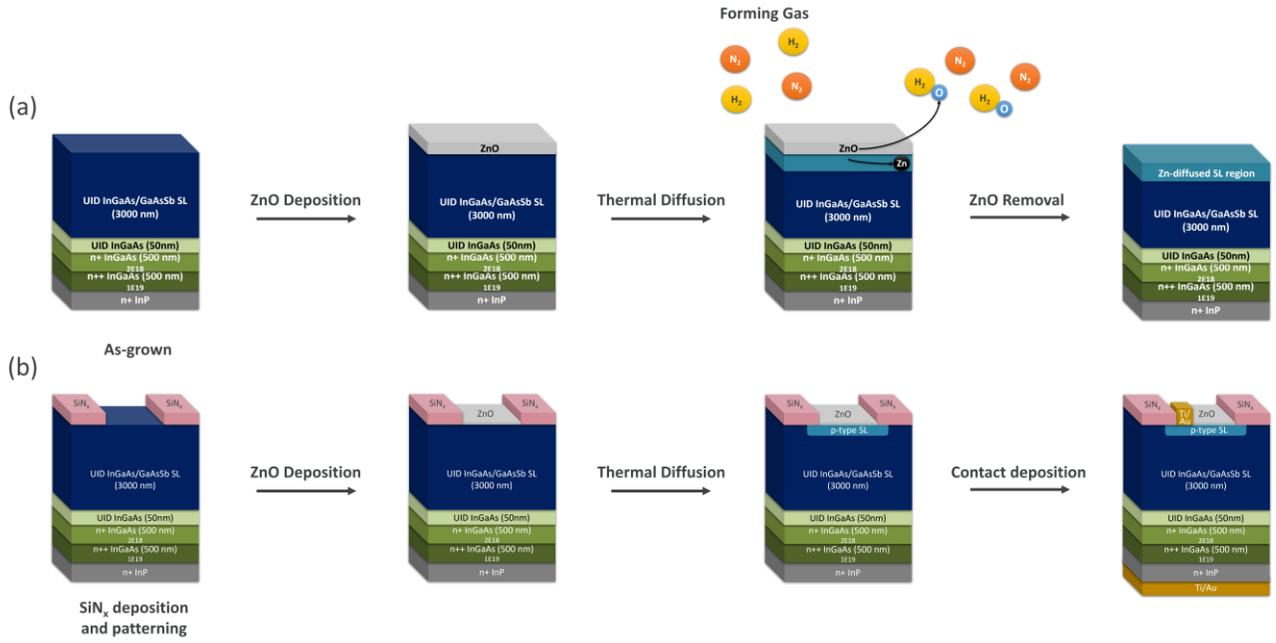

FIG. 1. (a) Schematic of the diffusion process using the as-grown sample and atomic layer deposition (b) Fabrication process for the planar Diodes

TABLE I. Different samples and their conditions for diffusion of Zn using ALD deposited ZnO. Sample A is the as-grown sample annealed at 450 °C.

| Sample name | ZnO thickness (nm) | Diffusion time (hour) |
|---|---|---|
| As-grown | 0 | 0 |
| A | 0 | 1 |
| B | 5 | 1 |
| C | 10 | 1 |
| D | 20 | 1 |
| E | 30 | 1 |

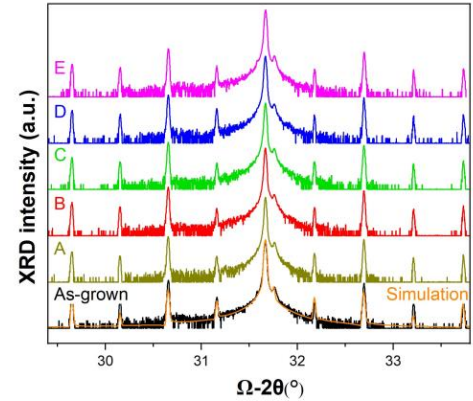

FIG. 2. X-ray diffraction (XRD) curves of the as-grown and the diffused superlattices with different Zn diffused conditions mentioned in TABLE I

Fabrication of the diodes is similar to the previously reported process[18]. To make a hard mask, $SiN_x$ was deposited on the UID InGaAs/GaAsSb SL using plasma etched chemical vapor deposition (PECVD). $SiN_x$ was then patterned using lithography and using reactive ion etching (RIE). ZnO was then deposited on top of the patterned $SiN_x$ hard mask and then thermally annealed for selectively diffusing Zn. After diffusion, ZnO was wet etched to deposit Ti/Au metal contacts. The schematic of the fabrication process is shown in FIG. 1 (b).

After the completion of the diffusion process, the effect of ZnO thickness and annealing was analyzed using high resolution x-ray diffraction (XRD). Each sample listed in TABLE I was measured using coupled omega-2theta scans on a Bruker D8 Discover system. The XRD measurement results, shown in FIG. 2, display no noticeable difference between the as-grown sample and the annealed or diffused samples. Further numerical analysis confirmed that the diffusion process caused no significant change to the full width half maximum (FWHM), position, or intensity of any of the peaks, indicating neither the lattice nor SL was affected by the process.

For all the samples, the zeroth order SL peak coincides with the substrate peak, indicating the average composition of the SL is lattice-matched to the InP substrate. Simulations using Bruker's LEPTOS 7.10 software indicate the small peak immediately to the right of the substrate peak is a reproduction of the substrate peak from the CuK$\alpha_2$ x-ray wavelength which is not fully eliminated by the equipment's 2-bounce monochromator. The thickness of the SL period, $\Lambda$, can be extracted from the spacing of SL satellite peaks using the equation. (2), derived from Bragg's law, where $\lambda$ is the x-ray wavelength, $\vartheta_n$ and $\vartheta_{SL}$ are the absolute angle of the SL satellite peak of order n and the zeroth order peak, respectively[20]. Using equation (2) the period of the SL was determined to be 10.18 nm, which was confirmed through simulation. Usually, the even ordered satellite peaks would be absent from the XRD of SL samples with 1:1 constitute thicknesses[20], however, our measurements showed the even ordered peaks to have a higher intensity. One possibility is these even order peaks originate from the presence of interfacial layers between the InGaAs and the GaAsSb layers. This hypothesis is supported by our simulation results, in which the even-order satellite peaks were absent unless an interfacial layer was added after either material. However, the observed behavior of higher intensity even ordered satellite peaks was only replicated when interfacial layers were incorporated after both the InGaAs and GaAsSb SL layers.

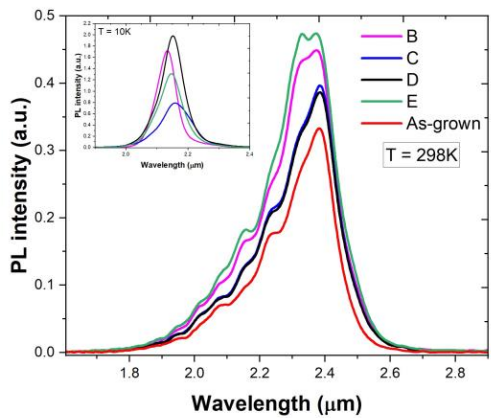

FIG. 3. PL spectra of the as-grown superlattice and Zn-diffused superlattices (B, C, D, E) at room temperature and (inset) PL measurements of the same samples at cooled temperature.

$$2\Lambda(sin\vartheta_n - sin\vartheta_{SL}) = \pm n\lambda \quad (2)$$

Photoluminescence (PL) spectra of Zn-diffused samples (B, C, D, E) and the as-grown samples were measured at different temperatures using a modulated continuous wave laser excitation (980 nm) with a power of 103.7 mW (FIG. 3). It is observed that PL intensity for the Zn-diffused samples increases slightly as compared to the as-grown sample, and the peak for all samples is obtained at 2.37 μm, corresponding to a bandgap of 0.52 eV. At room temperature, the left side of the PL peak has step-like features, which may suggest higher energy transitions. At cooled temperatures, the PL intensity increases and the PL peak blue shifts to 2.14 ± 0.01 μm due to the increase in the bandgap energy. Further analysis is needed to understand the photoluminescence behavior of the superlattice.

Post diffusion, Zn concentration, and diffusion depth were previously reported to be approximately $1\times10^{20} cm^{-3}$ and 50 nm, respectively[18]. To determine the background doping concentration, planar diodes were fabricated, and the capacitance-voltage (CV) measurements were conducted. FIG. 4 (a) shows the CV measurement of the fabricated planar diodes of different diameters (50 μm to 150 μm). As expected, the capacitance of the devices scaled with the area of the diodes. On further analysis, it was observed that the area-normalized capacitance of the planar diodes varied for varying device sizes. The variation in area-normalized capacitance indicated that the area of the devices was larger than the originally designed mask area. The additional area can be accounted for diffusion taking place in the lateral direction along with the vertical direction. Schematic of the phenomenon is shown in FIG. 4 (b) - inset, where $L_m$ is the initially designed mask diameter, and $L_d$ is the lateral diffusion.

TABLE II. The calculated diode diameter and the calculated lateral diffusion corresponding to the designed diameter are shown in the above table. The designed diode diameter is the diameter (of the device) in the SiN$_x$ hard mask before the deposition of ZnO.

| Designed diode diameter - $L_m$ (μm) | Calculated diode diameter $L_c$ (μm) | Calculated lateral diffusion - $L_d$ (μm) |
|---|---|---|
| 50 | 110 | 30 |
| 80 | 130 | 25 |
| 100 | 145 | 22.5 |
| 150 | 190 | 20 |

To analyze the lateral diffusion, we corrected the area of the planar diodes to match the doping profile of a previously reported InGaAs/GaAsSb SL mesa diode[9]. The doping profile of the fabricated planar diodes should match the doping profile of the reference mesa diode since the material in both cases is the same. FIG. 4 (b) shows the doping concentration vs the depletion width of the planar diodes with corrected areas and the reference mesa diode. FIG. 4 (b) shows that the diodes were fully depleted around 3000 nm, which agrees with the previously reported mesa diode[9] as well as initially grown superlattice structure. Finally, the background doping was determined to be $5.8×10^{14} cm^{-3}$.

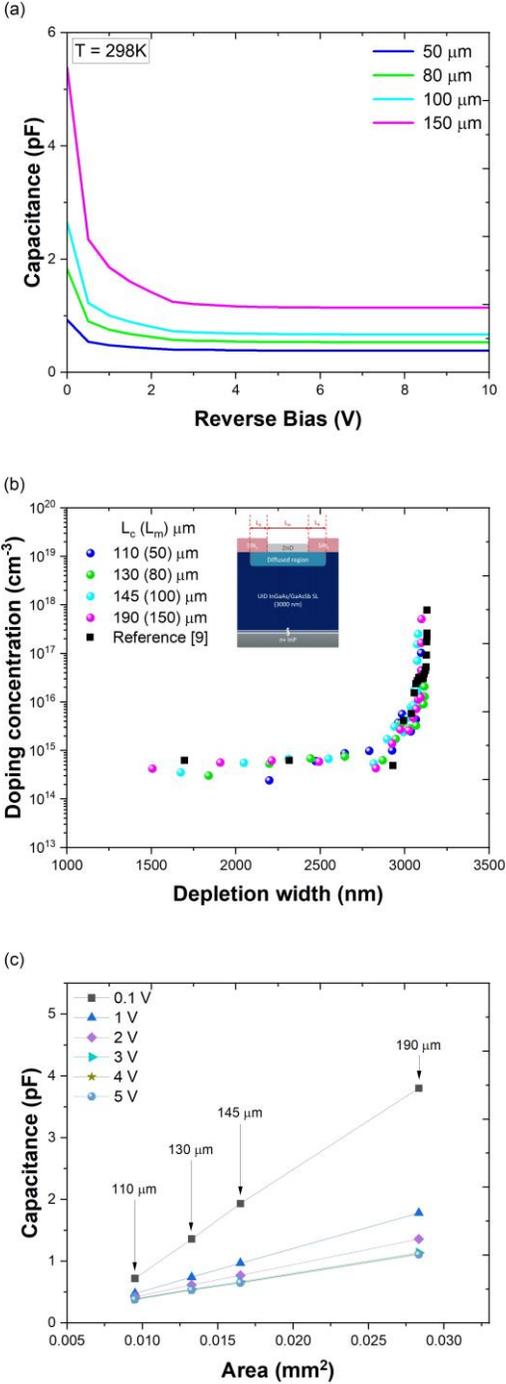

FIG. 4. Capacitance-voltage analysis (a) Capacitance-voltage (CV) measurement of the diodes of different sizes fabricated with sample D. (b) Doping concentration of the different diodes and their respective depletion width. Reference[9] diode is the mesa diode of 200 µm diameter. (Inset) Schematic showing the designed mask diameter ($L_m$ (µm)) and the lateral diffusion ($L_d$ (µm)) in the planar diodes. (c) Capacitance-Area analysis of the diodes at different voltages.

Further, to quantify the lateral diffusion, the corrected areas were calculated using the new diameter, which is $L_m+2L_d$. The lateral diffusion of Zn in each diode was calculated to be between 20 µm to 30 µm and is shown in Table II. The calculated lateral diffusion is longer for devices with smaller areas. When the SiNx mask is etched to selectively diffuse Zn, the SiNx mask around the etched region may be affected. The edges of the mask may become thinner, which may increase the diffusion under the mask. This phenomenon can be associated with the stress generated at the edges of the mask during the diffusion process and is prominent for diffusion in smaller-sized planar diodes[21]. However, little is known about the cause of high lateral diffusion, and more research is needed to understand the underlying physics. Finally, the corrected areas of the diodes are then used to plot the graph for the capacitance vs the corrected area FIG. (4 (c)). The graph shows that the capacitance scales linearly with the corrected area of diodes. The linear nature of the capacitance-area plot and FIG. 4 (c) shows evidence that the lateral diffusion of the planar diodes is accurately calculated.

Room temperature current-voltage (IV) measurements of the photodiodes of various diameters are shown in FIG. 5 (a). The rectifying nature of the IV curve depicts the formation of a p-i-n junction due to the diffusion of Zn into the UID superlattice. The lowest dark current obtained at room temperatures for -2V is 1 µA and it was obtained for the device with 30 µm designed diameter. The ideality factor for this diode was determined to be 2. The dark current density for the didoes was calculated using the corrected area and the graph is shown in the supplementary FIG. A 1 (b). From the current density curve, it can be inferred that the devices operate in the bulk-limited regime, and the surface current is suppressed. The obtained current density of the diodes is comparable to the reported InGaAs/GaAsSb SL diodes[9,16]. The photocurrent of the diode was measured using a 2 µm laser source with a 10 µm spot size and 70mW optical power. FIG. 5 (c) shows a higher photocurrent with the respective dark current of the diode.

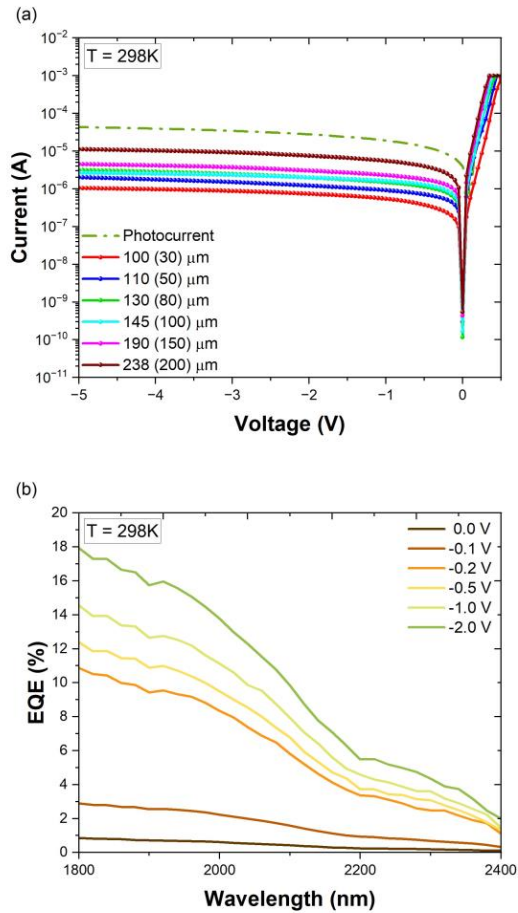

FIG. 5. (a) Current-voltage measurements of the diodes fabricated with sample D. The photocurrent was measured with 130 (80) μm diode. (b) Quantum Efficiency measurements of a diode with 238 μm diameter.

The external quantum efficiency (EQE) was calculated and plotted in FIG. 5 (b). The 50% cut-off wavelength was observed around 2150 nm and the full cutoff was observed beyond 2400 nm. FIG. 5 (b) shows the calculated EQE up to 2400 nm due to the limitation of the reference detector. The EQE was determined to be 11.11% at 2μm wavelength at an applied bias of -1V. The achieved EQE is lower than the previously reported EQE of similar materials[9,16]. More research is needed to understand the reason for the lower EQE of the fabricated planar diodes.

In conclusion, Zn-diffusion was demonstrated using an alternate method of using ZnO deposited through atomic layer deposition. The effect of Zn-diffusion on the superlattice material is studied using XRD, and PL. Material characterization showed that the diffusion process does not affect the superlattice, hence, maintaining the quality of the superlattice. The lateral diffusion in the planar diodes is analyzed and quantified using CV measurements. IV characteristics post-diffusion showed evidence of Zn diffusing into the superlattice, leading to the formation of a PIN diode. The QE measured at 2μm wavelength was 11.11%, and the reason for the low QE is still unknown. Further optimization of the diffusion process is required to reduce the dark current of the diodes.

ACKNOWLEDGMENTS

Dr. Charles Reyner, Dr. Gamini Ariyawansa, and Dr. Brent Webster from the Air Force Research Laboratory (AFRL), Dayton, OH, contributed to detailed discussions about the process, and some of the diffusion processes were completed at the AFRL laboratory. Funding: This work was supported by the Advanced Component Technology (ACT) program of NASA's Earth Science Technology Office (ESTO) under Grant No. 80NSSC21K0613.

1. M. B. Frish, "Current and emerging laser sensors for greenhouse gas sensing and leak detection," in *Next-Generation Spectroscopic Technologies VII*, Vol. 9101 (SPIE, 2014) pp. 122–133.
2. Z. Xie, Z. Deng, X. Zou, and B. Chen, "InP-based near infrared/extendedshort wave infrared dual-band photodetector," IEEE Photonics Technology Letters 32, 1003–1006 (2020).
3. J. Wang, Z. Xie, L. Zhu, X. Zou, X. Zhao, W. Yu, R. Liu, A. Du, Q. Gong, and B. Chen, "InP-Based Broadband Photodetectors With InGaAs/GaAsSb Type-II Superlattice," IEEE Electron Device Letters 43, 757–760 (2022).
4. Y. Chen, X. Zhao, J. Huang, Z. Deng, C. Cao, Q. Gong, and B. Chen, "Dynamic model and bandwidth characterization of InGaAs/GaAsSb type-II quantum wells PIN photodiodes," Optics express 26, 35034–35045 (2018). [6]D. Wu, A. Dehzangi, J. Li, and M. Razeghi, "High performance Zndiffused planar mid-wavelength infrared type-II InAs/InAs$_{1-x}$Sb$_x$ superlattice photodetector by MOCVD," Applied Physics Letters 116 (2020).
5. M. Muduli, M. Schwartz, N. Gajowski, S. Lee, and S. Krishna, "Investigation of Zn-diffusion in 2-micron InGaAs/GaAsSb superlattice planar diodes using atomic layer deposition of ZnO," in *Infrared Technology and Applications XLIX*, Vol. 12534 (SPIE, 2023) pp. 45–50.
6. G. Ariyawansa, J. M. Duran, C. J. Reyner, and J. E. Scheihing, "Surface passivation for PhotoDetector applications," (2019), US Patent 10,297,708. [20]D. K. Bowen and B. K. Tanner, *High resolution X-ray diffractometry and topography* (CRC press, 1998).
7. K. Shih and J. Blum, "Al$_x$Ga$_{1-x}$As Grown-Diffused Electroluminescent Planar Monolithic Diodes," Journal of Applied Physics 43, 3094–3097 (1972).
8. [1]A. Joshi and D. Becker, "High-Speed Low-Noise p-i-n InGaAs Photoreceiver at 2μm Wavelength," IEEE Photonics Technology Letters 20, 551–553 (2008).
9. [7]E. A. Plis, "InAs/GaSb type-II superlattice detectors," Advances in Electronics 2014 (2014).
10. [8]B. Marozas, W. Hughes, X. Du, D. Sidor, G. Savich, and G. Wicks, "Surface dark current mechanisms in III-V infrared photodetectors," Optical Materials Express 8, 1419–1424 (2018).
11. "Growth and characterization of InGaAs/GaAsSb type II superlattice absorbers for 2 μm avalanche photodiodes, author=Jung, Hyemin and Lee, Seunghyun and Schwartz, Mariah and Guo, Bingtian and Grein, Christoph H and Campbell, Joe C and Krishna, Sanjay," in *Infrared Technology and Applications XLVIII*, Vol. 12107 (SPIE, 2022) pp. 96–104.
12. E. A. Plis, M. N. Kutty, and S. Krishna, "Passivation techniques for InAs/GaSb strained layer superlattice detectors," Laser & Photonics Reviews 7, 45–59 (2013).
13. O. d. M. Braga, C. A. Delfino, R. M. S. Kawabata, L. D. Pinto, G. S. Vieira, M. P. Pires, P. L. d. Souza, E. Marega, J. A. Carlin, and S. Krishna, "Investigation of ingaas/inp photodiode surface passivation using epitaxial regrowth of inp via photoluminescence and photocurrent," Materials Science in Semiconductor Processing 154, 107200 (2023).
14. [11]R. K. Saroj, S. Slivken, G. J. Brown, M. Razeghi, *et al.*, "Demonstration of Zn-diffused planar long-wavelength infrared photodetector based on typeII superlattice grown by MBE," IEEE Journal of Quantum Electronics 58, 1–6 (2022).
15. [12]A. Dehzangi, D. Wu, R. McClintock, J. Li, and M. Razeghi, "Planar nBn type-II superlattice mid-wavelength infrared photodetectors using zinc ionimplantation," Applied Physics Letters 116 (2020).
16. [13]A. Dehzangi, D. Wu, R. McClintock, J. Li, A. Jaud, and M. Razeghi, "Demonstration of planar type-II superlattice-based photodetectors using silicon ion-implantation," in *Photonics*, Vol. 7 (MDPI, 2020) p. 68.
17. [4567]H. Inada, H. Mori, Y. Nagai, Y. Iguchi, T. Saitoh, K. Fujii, T. Ishizuka, and K. Akita, "MOVPE grown InGaAs/GaAsSb type II quantum well photodiode for SWIR focal plane array," in *Infrared Technology and Applications XXXVII*, Vol. 8012 (SPIE, 2011) pp. 648–653.
18. [16]Y. Uliel, D. Cohen-Elias, N. Sicron, I. Grimberg, N. Snapi, Y. Paltiel, and M. Katz, "InGaAs/GaAsSb Type-II superlattice based photodiodes for short wave infrared detection," Infrared Physics & Technology 84, 63–71 (2017). [17]N. H. Ky, L. Pavesi, D. Araujo, J. Ganiere, and F. Reinhart, "A model for the Zn diffusion in GaAs by a photoluminescence study," Journal of applied physics 69, 7585–7593 (1991).